\documentclass[aps,preprint]{revtex4}%
\usepackage{amsfonts}
\usepackage{amsmath}
\usepackage{amssymb}
\usepackage{graphicx}%
\providecommand{\U}[1]{\protect\rule{.1in}{.1in}}
\newtheorem{theorem}{Theorem}
\newtheorem{acknowledgement}[theorem]{Acknowledgement}

\begin{document}
\preprint{ }
\title[Higgs for Graviton]{Higgs for Graviton: Simple and Elegant Solution}
\author{Ali H.Chamseddine}
\affiliation{American University of Beirut, Physics Department, Beirut, Lebanon, and
I.H.E.S. F-91440 Bures-sur-Yvette, France}
\author{Viatcheslav Mukhanov}
\affiliation{Theoretical Physics, Ludwig Maxmillians University,Theresienstr. 37, 80333
Munich, Germany and Department of Physics, New York University, NY 10003, USA}
\pacs{PACS number}

\begin{abstract}
A\ Higgs mechanism for gravity is presented, where four scalars with global
Lorentz symmetry are employed. We show that in the \textquotedblleft broken
symmetry phase\textquotedblright\ a graviton absorbs all scalars and become
massive spin 2 particle with five degrees of freedom. The resulting theory is
unitary and free of ghosts.

\end{abstract}
\volumeyear{ }
\volumenumber{ }
\issuenumber{ }
\eid{ }
\startpage{1}
\endpage{7}
\maketitle





\section{ Introduction}

Forty years ago van Dam, Veltman \cite{vanDam} and Zakharov \cite{Zakharov}
pointed out that the propagator for a massive graviton does not have a smooth
limit to the massless case. The action used is that of Fierz and Pauli
\cite{Fierz} with mass terms breaking general coordinate invariance
explicitly. The straightforward conclusion was that the graviton mass must be
mathematically strictly zero rather than some extremely small value because in
the presence of discontinuity the massless and massive theories would predict
different results either for the perihelion shift or deflection of starlight.
This apparent paradox was resolved by Vainstein \cite{Vainshtein} who found
that the massive theory contains a new distance scale below which the massive
graviton behave like massless particle and it became clear that the graviton
could have small nonvanishing mass which still would not contradict
experiments. Over the years, further development of this scale were considered
and in \cite{Deffayet} it was clearly demonstrated how this mechanism works
(see, also \cite{Babichev}).

The analysis by Deser and Boulware \cite{Boulware} lead to the conclusion that
the massive theory is ill behaved because in addition to the five degrees of
freedom of massive graviton there must be an extra scalar degree of freedom,
which does not decouple. Work by Isham, Salam and Strathdee \cite{Isham}
examined a theory of bigravity with a direct mixing mass term, where one of
the gravitons becomes massive while the other remained massless. This was
generalized by Chamseddine, Salam and Strathdee \cite{Chamseddine}, who
considered the mixing mass terms generated through the spontaneous breaking of
gauge symmetry. (for further developments in bigravity theories in relation
with the graviton mass see \cite{ArkaniHamed} and references there). There
were also attempts to use theories with extra dimensions. Dvali, Gabadadze,
and Porrati \cite{Dvali} have invented a model based on five dimensions with
an infinite size extra dimension. Their theory when considered around a true
background seems to be free of ghosts. This theory is especially interesting
because of the claim of uniqueness \cite{Hofmann}. Further interesting steps
were made in \cite{Gabadadze}, \cite{deRham}, where general relativity with an
auxiliary non-dynamical extra dimension was considered with the purpose of
obtaining effective massive ghost-free gravity.

It was suspected that the failure of obtaining a ghost free consistent theory
for a massive graviton in four dimensions with only one metric, is related to
the absence of a ghost free Higgs mechanism that would generate the graviton
mass. The string inspired theories, considered in \cite{Green}, \cite{Siegel}
are not ghost free when considered around trivial background.

In a promising attempt 't Hooft \cite{'tHooft} (see also \cite{Kakushadze3},
\cite{Demir}) exploited a collection of four scalar fields whose vacuum
expectation value breaks general coordinate invariance to give mass to the
graviton. The kinetic energies of the scalar fields were combined together
using the Minkowski metric and thus involving a ghost in the unbroken phase.
In the broken symmetry phase the model failed to produce the Fierz-Pauli term
for the massive graviton, and the ghost state could not be decoupled.

In this letter we give an elegant solution to the problem of making the
graviton massive via Higgs mechanism, and show explicitly how all Higgs fields
are absorbed. The resulting spectrum in the broken symmetry phase will consist
only of a massive graviton, with Fierz-Pauli mass term and, hence, it has five
degrees of freedom. In the unbroken phase we have a massless graviton
interacting with four scalar fields, which in the linear approximation lack a
propagator. The resulting theory is well defined in all different vacua and is
ghost free.

Let us consider four fields $\phi^{A}$, $A=0,1,2,3,$ which are scalars under
coordinate transformations and assume that they posses an extra symmetry with
respect to \textquotedblleft\emph{Lorentz transformations\textquotedblright}
in the field space. These transformations involve index $A,$ thus mixing the
scalar fields \ and preserving the metric $\eta_{AB}=\mathrm{diag}\left(
1,-1,-1,-1\right)  $ in the field space. Next, from $\phi^{A}$ we construct
the field space tensor%
\begin{equation}
H^{AB}=g^{\mu\nu}\partial_{\mu}\phi^{A}\partial_{\nu}\phi^{B},
\end{equation}
symmetric with respect to $A$ \ and $B.$ The scalar field indices $A$ and $B$
will always be raised and lowered with Minkowski metric $\eta_{AB}.$ It is
convenient to decompose $H_{\,B}^{A}$ into trace and traceless parts as
\begin{equation}
H_{\,B}^{A}=\tilde{H}_{\,B}^{A}+\frac{1}{4}\delta_{B}^{A}H, \label{1a}%
\end{equation}
where $H=H_{A}^{A}$ and $\tilde{H}_{\,A}^{A}=0.$

To demonstrate the idea we will first consider the following action which is
explicitly diffeomorphism and Lorentz invariant and provides us the graviton
mass term :
\begin{equation}
S=-\frac{1}{2}%
{\displaystyle\int}
d^{4}x\sqrt{-g}R+\frac{m^{2}}{2}\int d^{4}x\sqrt{-g}\left[  3\left(  \left(
\frac{1}{4}H\right)  ^{2}-1\right)  ^{2}-\ \tilde{H}_{\,B}^{A}\tilde{H}%
_{\,A}^{B}\ \right]  , \label{1}%
\end{equation}
where $8\pi G=1.$ It is easy to see that the equations of motion for the
metric $g_{\mu\nu}$ and fields $\phi^{A}$ admit the following vacuum Minkowski
solution
\begin{equation}
\left\langle g_{\mu\nu}\right\rangle =\eta_{\mu\nu},\quad\phi^{A}=x^{A}.
\label{2}%
\end{equation}
It is this solution that \emph{identifies} the global Minkowski metric
$\eta_{AB}$ with that of space-time $\eta_{\mu\nu}.$ We now expand the fields
around this vacuum
\begin{equation}
\phi^{A}=x^{A}+\chi^{A},\quad\text{ }g^{\mu\nu}=\eta^{\mu\nu}+h^{\mu\nu},
\label{3}%
\end{equation}
Introducing
\begin{equation}
\bar{h}_{\,}^{AB}=H^{AB}-\eta^{AB}=h^{AB}+\partial^{A}\chi^{B}+\partial
^{B}\chi^{A}+h^{AC}\partial_{C}\chi^{B}+h^{BC}\partial_{C}\chi^{A}%
+h^{CD}\partial_{C}\chi^{A}\partial_{D}\chi^{B}, \label{3a}%
\end{equation}
where $h^{AB}=h^{\mu\nu}\delta_{\mu}^{A}\delta_{\nu}^{B}$ and $\partial
^{A}=\delta_{\mu}^{A}\eta^{\mu\nu}\partial_{\nu}$, we can rewrite action
(\ref{1}) in the following form%
\begin{equation}
S=-\frac{1}{2}%
{\displaystyle\int}
d^{4}x\sqrt{-g}R+\frac{m^{2}}{2}\int d^{4}x\sqrt{-g}\left[  \left(  \bar
{h}^{2}-\bar{h}_{B}^{A}\bar{h}_{A}^{B}\right)  +\frac{3}{4^{2}}\bar{h}%
^{3}\ +\frac{3}{4^{4}}\bar{h}^{4}\right]  . \label{3b}%
\end{equation}
where $\bar{h}_{B}^{A}=\eta_{BC}h^{AC},$ $\bar{h}=\bar{h}_{A}^{A}.$ Note that
this result is exact and we did not use any approximation to derive it.
\ Moreover the variable $\bar{h}_{\,B}^{A}$ is diffeomorphism invariant up to
an arbitrary order in perturbations.

Now let us consider small perturbations around background (\ref{2}). Then up
to the linear order in perturbations $\chi^{A}$ and $h^{\mu\nu}$,
\begin{equation}
\bar{h}_{\,B}^{A}=h_{\,B}^{A}+\partial_{B}\chi^{A}+\partial^{A}\chi_{B},
\label{5}%
\end{equation}
Einstein action is invariant under infinitesimal transformations $\tilde
{x}=x+\xi,$ where metric perturbations around Minkowski space-time transform
in a way similar to (\ref{5}), with $\chi$ replaced by $\xi.$ Therefore the
full action, up to second order terms could be expressed in terms of $\bar
{h}_{B}^{A}:$%
\begin{equation}
S=\frac{1}{2}%
{\displaystyle\int}
d^{4}x\left[  \bar{h}_{B}^{A,C}\bar{h}_{A,C}^{B}-2\bar{h}_{C}^{A,C}\bar
{h}_{A,D}^{\,\,D}+2\bar{h}_{C}^{A,C}\bar{h}_{,A}-\bar{h}_{,A}\bar{h}%
^{,A}-m^{2}\left(  \bar{h}_{B}^{A}\bar{h}_{A}^{B}-\bar{h}^{2}\right)  \right]
. \label{9}%
\end{equation}
This clearly shows that the Higgs fields $\phi^{A}$ are completely absorbed to
form the massive graviton with five degrees of freedom described by
Fierz-Pauli mass term. Because we have avoided to include a term linear in $H$
in action (\ref{1}) the theory is free of ghosts even around a background with
$H_{B}^{A}=0.$

One can wonder how four degrees of freedom for the scalar fields (expected
naively) could disappear giving only three extra degrees of freedom to the
graviton. To understand this let us take the limit of vanishing gravitational
constant. In this case we must set $h_{B}^{A}=0$ in equation (\ref{5}), which
then becomes
\[
\bar{h}_{\,B}^{A}=\partial_{B}\chi^{A}+\partial^{A}\chi_{B},
\]
and in the action
\begin{equation}
\frac{1}{2}\int d^{4}x\left(  \bar{h}^{2}-\bar{h}_{B}^{A}\bar{h}_{A}%
^{B}\right)  =\int d^{4}x\left[  \left(  \partial_{A}\chi^{A}\right)
^{2}-\left(  \partial_{A}\chi^{B}\right)  \left(  \partial^{A}\chi_{B}\right)
\right]  ,
\end{equation}
one can immediately recognize Maxwell action for \textquotedblleft4-vector
potential\textquotedblright\ $\chi^{A}.$ Thus, around background (\ref{2}) the
perturbations of four scalar fields would lose one degree of freedom, the
$\chi^{0}$, to leave three independent physical degrees of freedom.

The action used is not the most general one. In fact, there exist infinitely
many actions which could serve the same purpose. This is not surprising
because even in the standard electroweak theory the uniqueness of the Higgs
potential is entirely due to the requirement of renormalizability of the
theory. In grand unified theories there are many possible choices for the
Higgs potential. Our action (\ref{1}) possess shift symmetry $\phi
^{A}\rightarrow\phi^{A}+c^{A}$ where $c^{A}$ are constants, and extra discrete
symmetry $H_{B}^{A}\rightarrow-H_{B}^{A}$. However, even these symmetries,
which could protect against the appearance of unwanted quantum corrections,
are not enough to fix the action unambiguously.

At first glance a possible simple action which could serve the purpose is
\begin{equation}
\frac{m^{2}}{2}\int d^{4}x\sqrt{-g}\left(  \bar{h}^{2}-\bar{h}_{B}^{A}\bar
{h}_{A}^{B}\right)  ,\label{10a}%
\end{equation}
which, when rewritten in terms of $H_{B}^{A}=\delta_{B}^{A}+\bar{h}_{B}^{A},$
takes the form%
\begin{equation}
\frac{m^{2}}{2}\int d^{4}x\sqrt{-g}\left(  H^{2}-H_{B}^{A}H_{A}^{B}%
-6H+12\right)  .\label{11a}%
\end{equation}
In this form it is clear that the linear term contains a ghost. Nevertheless
this problem can be easily fixed by adding to the action terms which are not
\textquotedblleft Lorentz invariant\textquotedblright\ with respect to
transformations in the \textit{space of field configurations.} There is
nothing wrong with such terms because these still preserve diffeomorphism and
space-time Lorentz invariance. The only thing about which we have to take care
of is that the corresponding terms will not spoil the action in quadratic
order around Minkowski background (\ref{2}). For example, if we add to
(\ref{11a}) the term $2(H_{0}^{0}-1)^{3}$ the ghost disappears and around
Minkowski background the action (\ref{11a}) is modified to
\begin{equation}
\frac{m^{2}}{2}\int d^{4}x\sqrt{-g}\left(  \bar{h}^{2}-\bar{h}_{B}^{A}\bar
{h}_{A}^{B}+2\left(  \bar{h}_{0}^{0}\right)  ^{3}\right)  .\label{12a}%
\end{equation}
The last term here looks like a Lorentz violating term. However as we have
stressed above this does not mean that we have abandoned the fundamental
Lorentz invariance of space-time. Note that around a trivial background with
$H_{B}^{A}=0$ the linearized scalar fields are propagating and have three
degree of freedom.

Returning back to action (\ref{1}) we find that the trace of the energy
momentum of the scalar fields is equal to%
\begin{equation}
T_{\mu}^{\mu}=\frac{m^{2}}{2}\left(  \frac{3}{128}H^{4}-6\right)  ,
\end{equation}
and therefore energy is bounded from below. This action is ghost free in
linear order around both, trivial and Minkowski, backgrounds. On the other
hand, because the time derivative of the fields appear in the action in the
combination%
\[
\left(  \dot{\phi}^{0}\right)  ^{2}-\left(  \dot{\phi}^{i}\right)  ^{2},
\]
one may worry that the phase space of $\left(  \dot{\phi}^{0}\right)  ^{2}$
might be unbounded and the problem with ghosts can reappear at the nonlinear
level. This problem can be easily solved by adding to the action terms which
depend only on $H_{0}^{0}$ and do not modify the action at quadratic order
around Minkowski background.

When the background scalar fields disappear, that is $H_{B}^{A}=0,$ the
graviton decouples from the scalar fields and becomes massless. In this case,
however, there appears negative cosmological constant of order $m^{2}$ and the
solution of the Einstein equations is anti de Sitter space. One can naturally
ask whether the appearance of a negative cosmological constant is an inherent
property needed for producing the graviton mass via Higgs mechanism? In fact,
it is not the case and we can easily find an action with zero or positive
cosmological constant. For example, let us consider
\begin{equation}
\frac{m^{2}}{2}\int d^{4}x\sqrt{-g}\left[  \left(  \left(  \frac{1}%
{4}H\right)  ^{2}\mathrm{\ }-1\right)  ^{2}\left(  \alpha\left(  \frac{1}%
{4}H\right)  ^{2}\mathrm{\ }-\beta\right)  -\ \tilde{H}_{\,B}^{A}\tilde
{H}_{\,A}^{B}\ \right]  . \label{10}%
\end{equation}
If the constants $\alpha$ and $\beta$ satisfy the condition $\alpha-\beta=3$
then this action provides the Fierz-Pauli term in the broken symmetry state.
In the unbroken phase with $H_{B}^{A}=0$ the above action reduces to the
action with only a cosmological constant $\Lambda=-\frac{1}{2}m^{2}\beta.$
Thus, taking $\alpha=3$ and $\beta=0$ we obtain that the cosmological constant
is zero in broken as well as in unbroken phase and hence Minkowski space-time
is the solution of Einstein equations in both cases. Another interesting
choice of parameters is $\alpha=2$ and $\beta=-1,$ corresponding to a positive
cosmological constant of order $m^{2}$ in the unbroken phase. In this case,
either the graviton has mass $m$ in broken symmetry phase, or has a vanishing
mass (in unbroken phase) with a cosmological constant of order $m^{2}.$ Let us
take $m\sim H_{0},$ where $H_{0}$ is the value of the Hubble constant today.
Then we obtain that the theory under consideration inevitably leads either to
modification of gravity on Vainstein scale, which is $H_{0}^{-1},$ or to the
presence of a cosmological constant of order $H_{0}^{2}.$ This opens an
interesting possibility for interpretation of dark energy in the universe.

Finally we would like to know what is happening in the limit $m^{2}%
\rightarrow0.$ Let us take for definiteness action (\ref{1}). Redefining the
fields, $H_{B}^{A}\rightarrow\hat{H}_{B}^{A}=\sqrt{m}H_{B}^{A},$ and taking
the limit $m^{2}\rightarrow0$, then action (\ref{1}) reduces to%
\begin{equation}
-\frac{1}{2}%
{\displaystyle\int}
d^{4}x\sqrt{-g}R+\frac{3}{2}\int d^{4}x\sqrt{-g}\left(  \frac{1}{4}\hat
{H}\right)  ^{4}\ \label{11}%
\end{equation}
The broken symmetry phase with $\hat{H}=4$ corresponds to a huge negative
cosmological constant of the order of Planck value. Therefore it is clear that
the only solution is Minkowski space with $\hat{H}=0$ and massless graviton.

\begin{acknowledgement}
We are grateful to C. Deffayet, G. Dvali, , G. Gabadadze\ J. Garriga and W.
Siegel for helpful comments and discussions. The work of AHC is supported in
part by the National Science Foundation grant 0854779. V.M. is supported by
TRR 33 \textquotedblleft The Dark Universe\textquotedblright\ and the Cluster
of Excellence EXC 153 \textquotedblleft Origin and Structure of the
Universe\textquotedblright.
\end{acknowledgement}

\end{document}